\documentclass[letter]{aa}     
\usepackage{amssymb}
\usepackage{graphicx}
\usepackage{txfonts}           
\usepackage{amsmath}  

\begin{document}

   \title{The majority of hot Jupiters formed beyond the water ice line}

\author{
Yaxing He\inst{1,2}\corrauth{yaxinghe@sjtu.edu.cn}
\and
Bertram Bitsch\inst{2}
\and
Adrien Houge\inst{3}
\and
Joe Williams\inst{4}
\and
Masahiro Ogihara\inst{1,5}
}

\institute{
State Key Laboratory of Dark Matter Physics, Tsung-Dao Lee Institute,
Shanghai Jiao Tong University, 1 Lisuo Road, Shanghai 201210, China
\and
Department of Physics, University College Cork, Cork, T12 R229, Ireland
\and
Center for Star and Planet Formation, Globe Institute, University of Copenhagen, {\O}ster Voldgade 5-7, 1350 Copenhagen, Denmark
\and
School of Physics and Astronomy, University of Exeter, Stocker Road, Exeter EX4 4QL, United Kingdom
\and
School of Physics and Astronomy, Shanghai Jiao Tong University,
800 Dongchuan Road, Shanghai 200240, China
}
\date{Received 26 May 2026 / Accepted 15 July 2026}

\abstract{Atmospheric compositions of giant exoplanets can retain information about their formation environments, as volatile species condense at different temperatures in protoplanetary discs. We investigated whether the atmospheric compositions of hot Jupiters can constrain their formation locations. We performed planet formation simulations using the {\tt ChemComp} code, including pebble drift, pebble and gas accretion, planet migration, stellar abundances, and two additional chemical processes: thermal decomposition of refractory organics and CO/CO$_2$ trapping in water ice. We applied this framework to nine observed hot Jupiter systems and compared the resulting atmospheric metallicities (C/H and O/H) with observational constraints. We found that the observed atmospheric abundances of the nine hot Jupiter systems can be reproduced by planets forming at different locations relative to the H$_2$O and CO$_2$ snowlines. Our results suggest that at least six of the nine systems are consistent with formation beyond the H$_2$O snowline. Combined with the observed orbital separations, eccentricities, and spin--orbit obliquities, these inferred formation locations indicate that many systems likely experienced dynamical scattering followed by tidal evolution. Atmospheric abundances, in combination with detailed orbital parameters, can provide a powerful diagnostic of the formation and migration histories of hot Jupiter systems, opening up avenues to understand the origin of giant planets in general.}


   \maketitle
   \nolinenumbers

\section{Introduction}
Hot Jupiters are gas giant planets with orbital periods of up to a few days, and their origins remain a key challenge in planet formation theory.
Because such massive planets are unlikely to form in situ at their present orbital separations, they are generally thought to originate at larger separations in protoplanetary discs and subsequently migrate inward. Several formation pathways have been proposed, including disc-driven migration \citep[e.g.][]{1996Natur.380..606L,2008ApJ...685..584I,2009A&A...501.1139M,2015A&A...582A.112B} and high-eccentricity migration followed by tidal circularisation \citep[e.g.][]{1996Sci...274..954R,2007ApJ...669.1298F,2012ApJ...751..119B,2018ApJ...856...37B,2018ARA&A..56..175D}. Distinguishing between these scenarios requires observational constraints that connect the present-day properties of hot Jupiters to their formation histories.

The atmospheric compositions of exoplanets provide an important probe of their formation environments. Because major volatile species such as H$_2$O, CO$_2$, CH$_4$, and CO condense at different temperatures in protoplanetary discs, their snowlines separate regions with distinct gas and solid compositions \citep{2011ApJ...743L..16O,2014prpl.conf..363P,2017MNRAS.469.3994B,2018A&A...613A..14E,2022ApJ...934...74M}. The evolution and radial transport of gas and dust further redistribute these volatiles, producing time-dependent chemical structures in the disc \citep{2015ApJ...815..109P,2017MNRAS.469.3994B,2018ApJ...864...78K,2021A&A...654A..71S,2023A&A...677L...7M}.

Planets forming at different radial locations relative to these snowlines are expected to accrete material with different elemental abundances, which can leave observable signatures in their atmospheres \citep{2011ApJ...743L..16O,2017MNRAS.469.3994B,2017MNRAS.469.4102M,2021A&A...654A..71S,2022A&A...665A.138B}. As a result, the elemental abundances of planetary atmospheres, in particular the C/O ratio as well as the elemental C/H and O/H ratios, may retain information about the location where a planet accreted most of its gas and solids \citep{2017MNRAS.469.3910C,2021ApJ...909...40T}. Atmospheric compositions can be used to constrain the formation locations of giant planets in protoplanetary discs.

Recent planet formation models that couple disc evolution with pebble accretion, gas accretion, and planetary migration have demonstrated that the disc composition can strongly influence the atmospheric compositions of giant planets \citep{2021A&A...654A..71S,2022A&A...665A.138B,2023A&A...679L...7D,2023A&A...679A..42S,2024MNRAS.535..171P}. In particular, radial drift and evaporation of icy pebbles can enrich the gas phase in volatile elements, producing a wide range of atmospheric abundances depending on the formation location and migration history of the planet \citep{2017MNRAS.469.3994B,2022A&A...665A.138B,2026A&A...707A.276G,2026A&A...706A..30O}. 
However, most previous studies have assumed solar elemental abundances for the host star and have explored only a limited number of planetary systems, which may not fully capture the diversity observed among hot Jupiter atmospheres.

In this work, we extended the planet formation framework of \citet{2022A&A...665A.138B} by adopting stellar abundances from the Hypatia catalogue \citep{2014AJ....148...54H}, including thermal decomposition of refractory organics \citep{2025A&A...699A.227H} and the trapping of CO and CO$_2$ in amorphous water ice \citep{2025MNRAS.544.3562W}, and applying the model to nine hot Jupiters. Using simulations that followed the growth and migration of these planets, we computed their atmospheric abundances and compared them with the observational constraints. This comparison allowed us to constrain the formation locations of nine hot Jupiters relative to the H$_2$O and CO$_2$ snowlines.

\section{Model}\label{model}

We used the \texttt{ChemComp} code \citep{2021A&A...654A..71S} to simulate the growth and migration of hot Jupiters in evolving protoplanetary discs. The model couples viscous disc evolution with pebble growth and drift using the two-population algorithm of \citet{2012A&A...539A.148B}, gas and pebble accretion \citep{2014prpl.conf..547J,2021MNRAS.501.2017N} onto the planet, and planetary migration   \citep{2011MNRAS.410..293P,2021MNRAS.501.2017N}. A detailed description of the framework was provided by \citet{2021A&A...654A..71S}.

Compared to the baseline model of \citet{2022A&A...665A.138B}, we included two additional chemical processes that alter the radial distribution of volatile species in the disc. First, we accounted for the fact that carbon stored in refractory organics does not simply sublimate at the soot line like other volatile species, but instead irreversibly decomposes into gas-phase volatile carbon, represented by C$_2$H$_2$ in the model, in the inner disc \citep{2025A&A...699A.227H}.
Second, motivated by observational evidence for trapped volatiles in water-rich ices \citep{2024ApJ...975..166B}, we included the trapping of CO and CO$_2$ in amorphous water ice. The trapped CO and CO$_2$ are released when amorphous water ice crystallises at $130\,\mathrm{K}$, thereby modifying the gas-phase carbon and oxygen abundances in the inner disc \citep{2025MNRAS.544.3562W}. We show the combined effects of thermal decomposition and trapping on the C and O abundances within the disc in Appendix~\ref{appendixA}. 

In contrast to previous studies that assumed solar elemental abundances, we adopted host-star abundances from the Hypatia catalogue, a compilation of literature spectroscopic abundance measurements for stars within 150 pc of the Sun \citep{2014AJ....148...54H}, as shown in Appendix~\ref{appendixC}. These host-star abundances not only set the individual elemental abundances (e.g. C/H, O/H), but also the overall metallicity, which is a summation of the individual elements. Furthermore, instead of modelling only a small number of planets, we extended the framework to a sample of nine hot Jupiters \citep{2021Natur.598..580L,2021atat.confE..20P,2025AandA...699A.342B,2025NatAs...9..845E}. Six of these systems were taken from \citet{2025AandA...699A.342B}, while three additional systems (WASP-77A~b, $\tau$~Boötis~b, and WASP-121~b) were included to extend the sample to planets with available C/H and O/H constraints. These abundances were used as the main observational constraints in this work.

The formation efficiency of giant planets is influenced by the underlying disc properties, where more massive discs are more likely to form giant planets in the pebble accretion scenario \citep{2023A&A...679A..42S}. In addition, planetary embryos need to start accreting early in order to grow efficiently into giant planets, because embryos that start accreting at late times do not grow efficiently \citep{2023A&A...679A..42S}. Furthermore, \citet{2026A&A...707A.276G} found, using the same planetary formation model as \citet{2023A&A...679A..42S}, that the largest influence on planetary compositions originates from the planet's starting position and its migration velocity, which is set by the disc's viscosity. Considering that we only have two atmospheric constraints (C/H and O/H), we chose to vary only two parameters: the initial planetary embryo position and the disc viscosity, while leaving all other parameters fixed across all simulations.

For each system, we performed a suite of simulations covering a range of initial orbital separations, which were chosen to probe formation both inside and outside the major snowlines. The exact initial locations considered for each planet are shown in Fig.~\ref{fig:figB1}.
We varied the disc viscosity over $\alpha = 10^{-4}$, $5 \times 10^{-4}$, and $10^{-3}$, consistent with the values adopted in previous giant-planet formation studies using similar models \citep{2022A&A...665A.138B,2023A&A...679A..42S,2026A&A...707A.276G}, which also guided our choices of the remaining model parameters, such as the disc mass, disc radius, and fragmentation velocity.
We did not include multiple giant planets in the simulations that, in principle, could alter the disc composition and thereby the composition of the growing planets. However, \citet{2024A&A...691A..50E} showed that the formation of multiple giant planets does not influence their atmospheric composition significantly, because the planets form in a sequential way from inside out.

In our simulations, we followed the growth and migration of a single planetary embryo until it reached the mass of the corresponding observed hot Jupiter. We then recorded the atmospheric composition and orbital separation at this point for comparison with the observations. Continuing the calculation would require an additional prescription for limiting late-stage gas accretion; otherwise, the planet would generally continue to grow beyond the observed mass. We adopted a fixed disc mass of
$M_{\rm disc} = 0.07\,M_\star$ for all systems. We excluded simulations in which the planet reached its final mass before $1\,\mathrm{Myr}$, because protoplanetary discs typically survive for several Myr \citep{2001ApJ...553L.153H,2009AIPC.1158....3M,2011ARA&A..49...67W,2015A&A...576A..52R,2022ApJ...939L..10P}. The simulations affected by this criterion are discussed in Appendix~\ref{appendixB}.

We neglected planetesimal formation and accretion to avoid introducing additional unconstrained parameters. This was also motivated by \citet{2023A&A...679L...7D}, who showed that models with substantial planetesimal formation struggle to produce C/H and O/H ratios above a few times stellar, whereas larger enrichments can be produced in the pure pebble scenario \citep{2021A&A...654A..71S,2023A&A...679L...7D}.
All other model parameters were kept fixed and are summarised in Appendix~\ref{appendixC} and Table~\ref{tab:tableC1}.

\section{Results}\label{result}

\begin{figure}[ht!]
\centering
\includegraphics[width=\columnwidth]{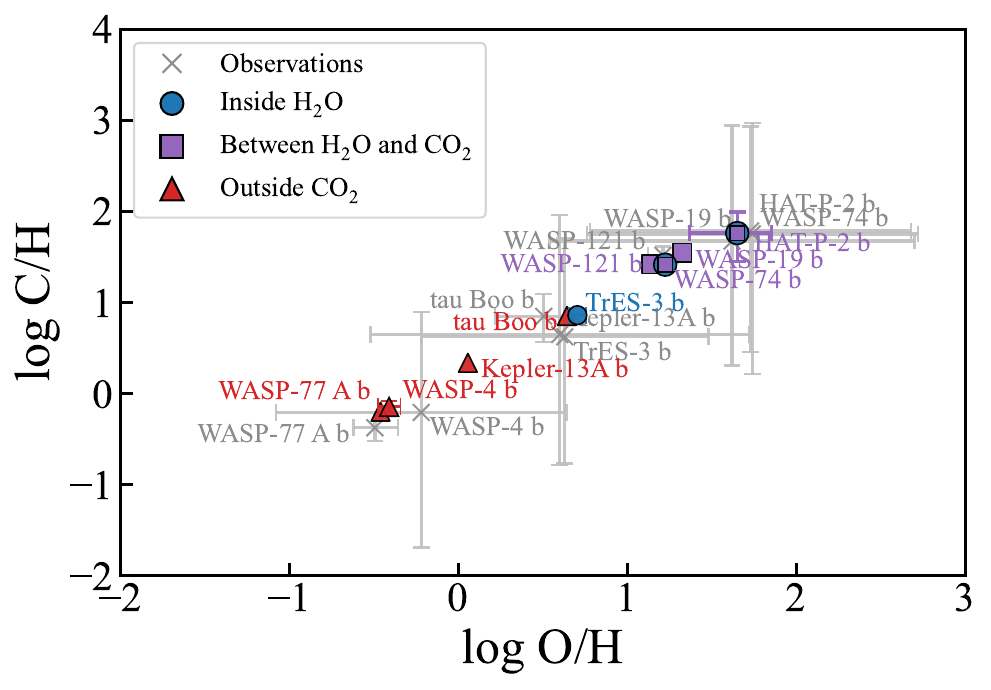}
\caption{
Simulated versus observed atmospheric metallicities of hot Jupiters in the C/H--O/H plane. Grey crosses show retrieved values listed in Table~\ref{tab:tableC1}(b); coloured symbols mark matching simulations. Both colour and marker shape indicate the formation region: blue circles inside the H$_2$O snowline, purple squares between the H$_2$O and CO$_2$ snowlines, and red triangles beyond the CO$_2$ snowline. For HAT-P-2~b and WASP-74~b, overlapping blue circles and purple squares denote simulations consistent with formation in both the inner and intermediate regions.
}
\label{fig:fig1}
\end{figure}

Figure~\ref{fig:fig1} summarises how the simulated atmospheric abundances depend on the planets' initial orbital locations.
Planets forming inside the H$_2$O snowline tend to exhibit enhanced oxygen abundances, reflecting the accretion of oxygen-rich gas. Planets forming between the H$_2$O and CO$_2$ snowlines occupy an intermediate region in the C/H--O/H plane, while planets forming beyond the CO$_2$ snowline show comparatively lower oxygen abundances. A similar trend is seen for carbon abundances, with planets forming in the inner disc generally exhibiting higher C/H ratios due to the enrichment of carbon-bearing species in the gas phase. Overall, the simulated atmospheric metallicities (C/H and O/H) show good agreement with the retrieved abundances of the nine hot Jupiters.

Individual systems further illustrate the interplay between stellar abundances and formation location. TrES-3~b is inferred to form inside the H$_2$O snowline. Because its host star is sub-solar in metallicity (Table~\ref{tab:tableC1}), its absolute atmospheric metallicity remains modest even though the planet is enriched relative to its host star.
In addition, HAT-P-2~b and WASP-74~b can be reproduced by models both inside the H$_2$O snowline and between the H$_2$O and CO$_2$ snowlines, indicating a degeneracy in their inferred formation locations. This degeneracy arises because simulated atmospheric compositions from both formation regions overlap with the observational constraints (as shown in Fig.~\ref{fig:fig1} and Appendix~\ref{appendixB}). It likely reflects the combined effects of migration and gas accretion, which can reduce compositional differences between planets forming at different initial locations.

The inferred formation regions are indicated by the marker styles and colours in Fig.~\ref{fig:fig1}. Most systems are consistent with formation beyond the H$_2$O snowline, with a substantial fraction forming even outside the CO$_2$ snowline. These results indicate that hot Jupiters do not predominantly form in the inner disc, but instead originate over a wide range of orbital separations and subsequently migrate inward to their present-day close-in orbits through disc-driven migration, with additional dynamical interactions required in the majority of systems (as shown in Fig.~\ref{fig:figB2}).

\section{Discussion}\label{sec:discussion}

\begin{table*}[ht!]
\centering
\caption{Inferred primordial orbital states and observed obliquities and eccentricities of nine hot Jupiter systems.}
\label{tab:hot_jupiters}
\footnotesize
\renewcommand{\arraystretch}{1.25}
\setlength{\tabcolsep}{2.5pt}
\begin{tabular}{lcccccc}
\hline\hline
Name & Inferred & Measured & $e_{\rm inferred}$ & $e_{\rm measured}$ & $\tau_{\rm realign}$  & $\tau_{\rm circ}$  \\
     & primordial obliquity & obliquity &  &  &(Gyr)  &(Gyr)  \\
\hline
WASP-77A~b
& misalign & align & $\neq 0$ & $0.0135^{+0.0006}_{-0.0006}$
& $0.3$--$3$
& $8\times10^{-3}$ \\

$\tau$~Boötis~b
& misalign & unknown & $\neq 0$ & $0.011^{+0.006}_{-0.006}$
& ($\gtrsim 4$--$40$)$^{a}$
& $2$ \\

WASP-121~b
& misalign & misalign & $\neq 0$ & $0$
& ($\gtrsim 1$--$10$)$^{a}$
& $10^{-3}$ \\

TrES-3~b
& align & unknown & $0$ & $0$
& ($0.2$--$2$)
& ($4\times10^{-3}$) \\

Kepler-13A~b
& misalign & misalign & $\neq 0$ & $0.00064^{+0.00012}_{-0.00016}$
& ($\gg 5$)$^{b}$
& $0.1$ \\

HAT-P-2~b
& align/misalign & align & $\neq 0$ & $0.517^{+0.002}_{-0.002}$
& ($\gtrsim 8$--$80$)$^{a}$
& ($80$) \\

WASP-74~b
& align/misalign & align & $0$ / $\neq 0$ & $0$
& ($6$--$60$)
& ($2\times10^{-2}$) \\

WASP-4~b
& misalign & align & $\neq 0$ & $0.005^{+0.004}_{-0.004}$
& $0.4$--$4$
& $3\times10^{-3}$ \\

WASP-19~b
& misalign & align & $\neq 0$ & $0.013^{+0.014}_{-0.009}$
& $0.1$--$1$
& $2\times10^{-4}$ \\
\hline
\end{tabular}
\tablefoot{
Parentheses indicate that the predicted primordial configuration already matches the observed one, without requiring tidal evolution; the enclosed timescale is shown only as a consistency check. Observed obliquities are from the Transiting Extrasolar Planet Catalogue (TEPCat) \citep{2011MNRAS.417.2166S}, and eccentricities are from the NASA Exoplanet Archive. The tidal timescales are estimated as described in Appendix~\ref{appendixD}. 
$^{a}$These hosts are close to or above the nominal $T_{\rm eff}\simeq6250\,{\rm K}$ transition discussed in Appendix~\ref{appendixD}; the realignment timescales should be regarded as lower limits.
$^{b}$Kepler-13~A is an A-type host, for which the inertial-wave realignment mechanism is expected to be inefficient; the quoted value should be regarded as a lower limit.
}
\end{table*}

The formation locations inferred from atmospheric compositions constrain the dynamical evolution of hot Jupiter systems. Planets that originate at large orbital separations must migrate inward, and in some systems disc-driven migration alone is insufficient.

The orbital configurations listed in Table~\ref{tab:hot_jupiters} are qualitative primordial states inferred from the migration pathway, rather than numerical obliquities or eccentricities calculated by the formation model. Systems that reach the inner disc through smooth disc-driven migration are interpreted as initially aligned and nearly circular. Systems that remain outside the inner disc boundary and require additional dynamical evolution to reach their currently observed orbits are classified as initially eccentric and possibly misaligned before tidal evolution.
In some cases, the planets cannot reach their current locations through smooth disc-driven migration alone (Fig.~\ref{fig:figB2}), suggesting additional dynamical interactions such as planet--planet scattering \citep[e.g.][]{2008ApJ...686..603J,2012ApJ...751..119B}. Scattering during the gas-disc phase could allow gas damping to reduce the excited eccentricity \citep[e.g.][]{2012MNRAS.419..366M,2013MNRAS.431.3494L,2023A&A...674A.178B}. Its effect on atmospheric composition would depend on the timing of scattering and the amount of gas accreted afterwards. Quantifying this process would require modelling multi-planet dynamics, which is beyond the scope of this work.
Spin--orbit misalignment could also be produced by tilting of the protoplanetary disc due to gravitational perturbations from nearby stars \citep{2012Natur.491..418B}.
The calculated tidal damping timescales are listed in Table~\ref{tab:hot_jupiters} and represent fiducial, order-of-magnitude estimates, with the prescriptions given in Appendix~\ref{appendixD}. Similar comparisons between eccentricities, tidal circularisation timescales, and system ages have also been used to observationally constrain hot-Jupiter migration pathways \citep{2025AJ....170..299K}. 

For the systems WASP-77A~b, WASP-4~b, and WASP-19~b, the inferred migration pathways suggest initially eccentric and possibly misaligned orbits. Their estimated circularisation and realignment timescales are typically shorter than the system ages. These systems can evolve from initially misaligned and eccentric configurations into the observed aligned and circular orbits.

For systems orbiting hotter stars, including $\tau$~Boötis~b, WASP-121~b, Kepler-13A~b, HAT-P-2~b, and WASP-74~b, spin--orbit realignment is less efficient. For WASP-121~b and Kepler-13A~b, the estimated circularisation timescales are shorter than the corresponding realignment timescales, so their eccentricities can be damped while their primordial spin--orbit misalignments are preserved. Such differential tidal evolution is consistent with the observed misaligned but nearly circular configurations of WASP-121~b and Kepler-13A~b, and provides a testable prediction that $\tau$~Boötis~b remains misaligned. 
For HAT-P-2~b and WASP-74~b, our simulations allow both aligned and misaligned primordial states. Given their observed alignments and the long estimated realignment timescales, we suggest that these systems were likely primordially aligned rather than tidally realigned.

HAT-P-2~b is a particularly massive planet ($M_p \sim 9\,M_{\rm Jup}$) with a large eccentricity ($e \approx 0.5$). Its estimated circularisation timescale is much longer than the system age, consistent with the survival of its high eccentricity. Its large eccentricity may have been enhanced by planet--disc interactions after the planet opened a deep gap \citep{2001A&A...366..263P,2006A&A...447..369K,2013A&A...555A.124B}.

We infer that TrES-3~b could have formed in an aligned, nearly circular orbit through smooth disc-driven migration, in general agreement with the sub-solar host star metallicity \citep{2013ApJ...767L..24D,2018ApJ...856...37B}. Its currently unknown obliquity would provide an additional testable prediction for future observations.

More generally, the broad range of inferred formation locations suggests that planetary embryos in the outer disc would need to form relatively early to accrete pebbles and gas before the disc dispersal. The formation of planetary embryos in rings associated with ice lines \citep{2022NatAs...6..357I} might play an important role in the origin of giant planets.

\section{Conclusions}

We investigated the formation locations of nine hot Jupiters using atmospheric C/H and O/H abundances as tracers of their birth environments. Using planet formation simulations that included pebble and gas accretion, migration, host-star abundances, refractory-organic decomposition, and CO/CO$_2$ trapping in water ice, we reproduced the observed atmospheric abundances of all nine systems.

The inferred formation locations span a wide range of orbital separations, with most planets forming beyond the H$_2$O snowline and several consistent with formation beyond the CO$_2$ snowline. Combined with their present-day orbital separations, eccentricities, and obliquities, these results suggest that dynamical interactions followed by tidal evolution may have played an important role in shaping a substantial fraction of the hot Jupiters in our sample, while smooth disc-driven migration can explain the remaining systems. Atmospheric abundances provide an important link between disc chemistry, giant-planet formation, and the dynamical evolution of hot Jupiter systems.

\begin{acknowledgements}
We thank Fei Dai and Rafael Luque for helpful discussions. This work is supported by the National Natural Science Foundation of China (grant No.12273023).    
\end{acknowledgements}

\bibliographystyle{aa}
\bibliography{aa}

@ARTICLE{2001ApJ...553L.153H,
       author = {{Haisch}, Jr., Karl E. and {Lada}, Elizabeth A. and {Lada}, Charles J.},
        title = "{Disk Frequencies and Lifetimes in Young Clusters}",
      journal = {\apjl},
         year = 2001,
        month = jun,
       volume = {553},
       number = {2},
        pages = {L153-L156},
          doi = {10.1086/320685},
archivePrefix = {arXiv},
       eprint = {astro-ph/0104347},
 primaryClass = {astro-ph},
       adsurl = {https://ui.adsabs.harvard.edu/abs/2001ApJ...553L.153H}
}

@ARTICLE{2011ARA&A..49...67W,
       author = {{Williams}, Jonathan P. and {Cieza}, Lucas A.},
        title = "{Protoplanetary Disks and Their Evolution}",
      journal = {\araa},
         year = 2011,
        month = sep,
       volume = {49},
       number = {1},
        pages = {67-117},
          doi = {10.1146/annurev-astro-081710-102548},
archivePrefix = {arXiv},
       eprint = {1103.0556},
 primaryClass = {astro-ph.GA},
       adsurl = {https://ui.adsabs.harvard.edu/abs/2011ARA&A..49...67W}
}

@ARTICLE{2015A&A...576A..52R,
       author = {{Ribas}, {\'A}lvaro and {Bouy}, Herv{\'e} and {Mer{\'\i}n}, Bruno},
        title = "{Protoplanetary disk lifetimes vs. stellar mass and possible implications for giant planet populations}",
      journal = {\aap},
         year = 2015,
        month = apr,
       volume = {576},
          eid = {A52},
        pages = {A52},
          doi = {10.1051/0004-6361/201424846},
archivePrefix = {arXiv},
       eprint = {1502.00631},
 primaryClass = {astro-ph.SR},
       adsurl = {https://ui.adsabs.harvard.edu/abs/2015A&A...576A..52R}
}

@ARTICLE{2022NatAs...6..357I,
       author = {{Izidoro}, Andre and {Dasgupta}, Rajdeep and {Raymond}, Sean N. and {Deienno}, Rogerio and {Bitsch}, Bertram and {Isella}, Andrea},
        title = "{Planetesimal rings as the cause of the Solar System's planetary architecture}",
      journal = {Nature Astronomy},
         year = 2022,
        month = mar,
       volume = {6},
        pages = {357-366},
          doi = {10.1038/s41550-021-01557-z},
archivePrefix = {arXiv},
       eprint = {2112.15558},
 primaryClass = {astro-ph.EP},
       adsurl = {https://ui.adsabs.harvard.edu/abs/2022NatAs...6..357I}
}

@ARTICLE{1996Natur.380..606L,
       author = {{Lin}, D.~N.~C. and {Bodenheimer}, P. and {Richardson}, D.~C.},
        title = "{Orbital migration of the planetary companion of 51 Pegasi to its present location}",
      journal = {\nat},
         year = 1996,
        month = apr,
       volume = {380},
       number = {6575},
        pages = {606-607},
          doi = {10.1038/380606a0},
       adsurl = {https://ui.adsabs.harvard.edu/abs/1996Natur.380..606L}
}

@ARTICLE{2008ApJ...685..584I,
       author = {{Ida}, S. and {Lin}, D.~N.~C.},
        title = "{Toward a Deterministic Model of Planetary Formation. V. Accumulation Near the Ice Line and Super-Earths}",
      journal = {\apj},
         year = 2008,
        month = sep,
       volume = {685},
       number = {1},
        pages = {584-595},
          doi = {10.1086/590401},
       adsurl = {https://ui.adsabs.harvard.edu/abs/2008ApJ...685..584I}
}

@ARTICLE{2009A&A...501.1139M,
       author = {{Mordasini}, C. and {Alibert}, Y. and {Benz}, W.},
        title = "{Extrasolar planet population synthesis. I. Method, formation tracks, and mass-distance distribution}",
      journal = {\aap},
         year = 2009,
        month = jul,
       volume = {501},
       number = {3},
        pages = {1139-1160},
          doi = {10.1051/0004-6361/200810301},
archivePrefix = {arXiv},
       eprint = {0904.2524},
 primaryClass = {astro-ph.EP},
       adsurl = {https://ui.adsabs.harvard.edu/abs/2009A&A...501.1139M}
}

@ARTICLE{2015A&A...582A.112B,
       author = {{Bitsch}, Bertram and {Lambrechts}, Michiel and {Johansen}, Anders},
        title = "{The growth of planets by pebble accretion in evolving protoplanetary discs}",
      journal = {\aap},
         year = 2015,
        month = oct,
       volume = {582},
          eid = {A112},
        pages = {A112},
          doi = {10.1051/0004-6361/201526463},
archivePrefix = {arXiv},
       eprint = {1507.05209},
 primaryClass = {astro-ph.EP},
       adsurl = {https://ui.adsabs.harvard.edu/abs/2015A&A...582A.112B}
}

@ARTICLE{1996Sci...274..954R,
       author = {{Rasio}, Frederic A. and {Ford}, Eric B.},
        title = "{Dynamical instabilities and the formation of extrasolar planetary systems}",
      journal = {Science},
         year = 1996,
        month = nov,
       volume = {274},
        pages = {954-956},
          doi = {10.1126/science.274.5289.954},
       adsurl = {https://ui.adsabs.harvard.edu/abs/1996Sci...274..954R}
}

@ARTICLE{2007ApJ...669.1298F,
       author = {{Fabrycky}, Daniel and {Tremaine}, Scott},
        title = "{Shrinking Binary and Planetary Orbits by Kozai Cycles with Tidal Friction}",
      journal = {\apj},
         year = 2007,
        month = nov,
       volume = {669},
       number = {2},
        pages = {1298-1315},
          doi = {10.1086/521702},
archivePrefix = {arXiv},
       eprint = {0705.4285},
 primaryClass = {astro-ph},
       adsurl = {https://ui.adsabs.harvard.edu/abs/2007ApJ...669.1298F}
}

@ARTICLE{2012ApJ...751..119B,
       author = {{Beaug{\'e}}, C. and {Nesvorn{\'y}}, D.},
        title = "{Multiple-planet Scattering and the Origin of Hot Jupiters}",
      journal = {\apj},
         year = 2012,
        month = jun,
       volume = {751},
       number = {2},
          eid = {119},
        pages = {119},
          doi = {10.1088/0004-637X/751/2/119},
archivePrefix = {arXiv},
       eprint = {1110.4392},
 primaryClass = {astro-ph.EP},
       adsurl = {https://ui.adsabs.harvard.edu/abs/2012ApJ...751..119B}
}

@ARTICLE{2018ARA&A..56..175D,
       author = {{Dawson}, Rebekah I. and {Johnson}, John Asher},
        title = "{Origins of Hot Jupiters}",
      journal = {\araa},
         year = 2018,
        month = sep,
       volume = {56},
        pages = {175-221},
          doi = {10.1146/annurev-astro-081817-051853},
archivePrefix = {arXiv},
       eprint = {1801.06117},
 primaryClass = {astro-ph.EP},
       adsurl = {https://ui.adsabs.harvard.edu/abs/2018ARA&A..56..175D}
}

@ARTICLE{2011ApJ...743L..16O,
       author = {{{\"O}berg}, Karin I. and {Murray-Clay}, Ruth and {Bergin}, Edwin A.},
        title = "{The Effects of Snowlines on C/O in Planetary Atmospheres}",
      journal = {\apjl},
         year = 2011,
        month = dec,
       volume = {743},
       number = {1},
          eid = {L16},
        pages = {L16},
          doi = {10.1088/2041-8205/743/1/L16},
archivePrefix = {arXiv},
       eprint = {1110.5567},
 primaryClass = {astro-ph.GA},
       adsurl = {https://ui.adsabs.harvard.edu/abs/2011ApJ...743L..16O}
}

@ARTICLE{2018A&A...613A..14E,
       author = {{Eistrup}, Christian and {Walsh}, Catherine and {van Dishoeck}, Ewine F.},
        title = "{Molecular abundances and C/O ratios in chemically evolving planet-forming disk midplanes}",
      journal = {\aap},
         year = 2018,
        month = may,
       volume = {613},
          eid = {A14},
        pages = {A14},
          doi = {10.1051/0004-6361/201731302},
archivePrefix = {arXiv},
       eprint = {1709.07863},
 primaryClass = {astro-ph.EP},
       adsurl = {https://ui.adsabs.harvard.edu/abs/2018A&A...613A..14E}
}

@ARTICLE{2022ApJ...934...74M,
       author = {{Molli{\`e}re}, Paul and {Molyarova}, Tamara and {Bitsch}, Bertram and {Henning}, Thomas and {Schneider}, Aaron and {Kreidberg}, Laura and {Eistrup}, Christian and {Burn}, Remo and {Nasedkin}, Evert and {Semenov}, Dmitry and {Mordasini}, Christoph and {Schlecker}, Martin and {Schwarz}, Kamber R. and {Lacour}, Sylvestre and {Nowak}, Mathias and {Schulik}, Matth{\"a}us},
        title = "{Interpreting the Atmospheric Composition of Exoplanets: Sensitivity to Planet Formation Assumptions}",
      journal = {\apj},
         year = 2022,
        month = jul,
       volume = {934},
       number = {1},
          eid = {74},
        pages = {74},
          doi = {10.3847/1538-4357/ac6a56},
archivePrefix = {arXiv},
       eprint = {2204.13714},
 primaryClass = {astro-ph.EP},
       adsurl = {https://ui.adsabs.harvard.edu/abs/2022ApJ...934...74M}
}

@ARTICLE{2017MNRAS.469.4102M,
       author = {{Madhusudhan}, Nikku and {Bitsch}, Bertram and {Johansen}, Anders and {Eriksson}, Linn},
        title = "{Atmospheric signatures of giant exoplanet formation by pebble accretion}",
      journal = {\mnras},
         year = 2017,
        month = aug,
       volume = {469},
       number = {4},
        pages = {4102-4115},
          doi = {10.1093/mnras/stx1139},
archivePrefix = {arXiv},
       eprint = {1611.03083},
 primaryClass = {astro-ph.EP},
       adsurl = {https://ui.adsabs.harvard.edu/abs/2017MNRAS.469.4102M}
}

@ARTICLE{2017MNRAS.469.3910C,
       author = {{Cridland}, A.~J. and {Pudritz}, Ralph E. and {Birnstiel}, Tilman and {Cleeves}, L. Ilsedore and {Bergin}, Edwin A.},
        title = "{Composition of early planetary atmospheres - II. Coupled Dust and chemical evolution in protoplanetary discs}",
      journal = {\mnras},
         year = 2017,
        month = aug,
       volume = {469},
       number = {4},
        pages = {3910-3927},
          doi = {10.1093/mnras/stx1069},
archivePrefix = {arXiv},
       eprint = {1705.02381},
 primaryClass = {astro-ph.EP},
       adsurl = {https://ui.adsabs.harvard.edu/abs/2017MNRAS.469.3910C}
}

@ARTICLE{2021A&A...654A..71S,
       author = {{Schneider}, Aaron David and {Bitsch}, Bertram},
        title = "{How drifting and evaporating pebbles shape giant planets. I. Heavy element content and atmospheric C/O}",
      journal = {\aap},
         year = 2021,
        month = oct,
       volume = {654},
          eid = {A71},
        pages = {A71},
          doi = {10.1051/0004-6361/202039640},
archivePrefix = {arXiv},
       eprint = {2105.13267},
 primaryClass = {astro-ph.EP},
       adsurl = {https://ui.adsabs.harvard.edu/abs/2021A&A...654A..71S}
}

@ARTICLE{2021ApJ...909...40T,
       author = {{Turrini}, D. and {Schisano}, E. and {Fonte}, S. and {Molinari}, S. and {Politi}, R. and {Fedele}, D. and {Pani{\'c}}, O. and {Kama}, M. and {Changeat}, Q. and {Tinetti}, G.},
        title = "{Tracing the Formation History of Giant Planets in Protoplanetary Disks with Carbon, Oxygen, Nitrogen, and Sulfur}",
      journal = {\apj},
         year = 2021,
        month = mar,
       volume = {909},
       number = {1},
          eid = {40},
        pages = {40},
          doi = {10.3847/1538-4357/abd6e5},
archivePrefix = {arXiv},
       eprint = {2012.14315},
 primaryClass = {astro-ph.EP},
       adsurl = {https://ui.adsabs.harvard.edu/abs/2021ApJ...909...40T}
}

@ARTICLE{2022A&A...665A.138B,
       author = {{Bitsch}, Bertram and {Schneider}, Aaron David and {Kreidberg}, Laura},
        title = "{How drifting and evaporating pebbles shape giant planets. III. The formation of WASP-77A b and {\ensuremath{\tau}} Bo{\"o}tis b}",
      journal = {\aap},
         year = 2022,
        month = sep,
       volume = {665},
          eid = {A138},
        pages = {A138},
          doi = {10.1051/0004-6361/202243345},
archivePrefix = {arXiv},
       eprint = {2207.06077},
 primaryClass = {astro-ph.EP},
       adsurl = {https://ui.adsabs.harvard.edu/abs/2022A&A...665A.138B}
}

@INPROCEEDINGS{2014prpl.conf..363P,
       author = {{Pontoppidan}, K.~M. and {Salyk}, C. and {Bergin}, E.~A. and {Brittain}, S. and {Marty}, B. and {Mousis}, O. and {{\"O}berg}, K.~I.},
        title = "{Volatiles in Protoplanetary Disks}",
    booktitle = {Protostars and Planets VI},
         year = 2014,
       editor = {{Beuther}, Henrik and {Klessen}, Ralf S. and {Dullemond}, Cornelis P. and {Henning}, Thomas},
        month = jan,
        pages = {363-385},
          doi = {10.2458/azu_uapress_9780816531240-ch016},
archivePrefix = {arXiv},
       eprint = {1401.2423},
 primaryClass = {astro-ph.EP},
       adsurl = {https://ui.adsabs.harvard.edu/abs/2014prpl.conf..363P}
}

@ARTICLE{2015ApJ...815..109P,
       author = {{Piso}, Ana-Maria A. and {{\"O}berg}, Karin I. and {Birnstiel}, Tilman and {Murray-Clay}, Ruth A.},
        title = "{C/O and Snowline Locations in Protoplanetary Disks: The Effect of Radial Drift and Viscous Gas Accretion}",
      journal = {\apj},
         year = 2015,
        month = dec,
       volume = {815},
       number = {2},
          eid = {109},
        pages = {109},
          doi = {10.1088/0004-637X/815/2/109},
archivePrefix = {arXiv},
       eprint = {1511.05563},
 primaryClass = {astro-ph.EP},
       adsurl = {https://ui.adsabs.harvard.edu/abs/2015ApJ...815..109P}
}

@ARTICLE{2017MNRAS.469.3994B,
       author = {{Booth}, Richard A. and {Clarke}, Cathie J. and {Madhusudhan}, Nikku and {Ilee}, John D.},
        title = "{Chemical enrichment of giant planets and discs due to pebble drift}",
      journal = {\mnras},
         year = 2017,
        month = aug,
       volume = {469},
       number = {4},
        pages = {3994-4011},
          doi = {10.1093/mnras/stx1103},
archivePrefix = {arXiv},
       eprint = {1705.03305},
 primaryClass = {astro-ph.EP},
       adsurl = {https://ui.adsabs.harvard.edu/abs/2017MNRAS.469.3994B}
}

@ARTICLE{2023A&A...679A..42S,
       author = {{Savvidou}, Sofia and {Bitsch}, Bertram},
        title = "{How to make giant planets via pebble accretion}",
      journal = {\aap},
         year = 2023,
        month = nov,
       volume = {679},
          eid = {A42},
        pages = {A42},
          doi = {10.1051/0004-6361/202245793},
archivePrefix = {arXiv},
       eprint = {2309.03807},
 primaryClass = {astro-ph.EP},
       adsurl = {https://ui.adsabs.harvard.edu/abs/2023A&A...679A..42S}
}

@ARTICLE{2023A&A...679L...7D,
       author = {{Danti}, C. and {Bitsch}, B. and {Mah}, J.},
        title = "{Composition of giant planets: The roles of pebbles and planetesimals}",
      journal = {\aap},
         year = 2023,
        month = nov,
       volume = {679},
          eid = {L7},
        pages = {L7},
          doi = {10.1051/0004-6361/202347501},
archivePrefix = {arXiv},
       eprint = {2310.02886},
 primaryClass = {astro-ph.EP},
       adsurl = {https://ui.adsabs.harvard.edu/abs/2023A&A...679L...7D}
}

@ARTICLE{2024MNRAS.535..171P,
       author = {{Penzlin}, Anna B.~T. and {Booth}, Richard A. and {Kirk}, James and {Owen}, James E. and {Ahrer}, E. and {Christie}, Duncan A. and {Claringbold}, Alastair B. and {Esparza-Borges}, Emma and {L{\'o}pez-Morales}, M. and {Mayne}, N.~J. and {McCormack}, Mason and {Meech}, Annabella and {Panwar}, Vatsal and {Powell}, Diana and {Sergeev}, Denis E. and {Taylor}, Jake and {Wheatley}, Peter J. and {Zamyatina}, Maria},
        title = "{BOWIE-ALIGN: how formation and migration histories of giant planets impact atmospheric compositions}",
      journal = {\mnras},
         year = 2024,
        month = nov,
       volume = {535},
       number = {1},
        pages = {171-186},
          doi = {10.1093/mnras/stae2362},
archivePrefix = {arXiv},
       eprint = {2407.03199},
 primaryClass = {astro-ph.EP},
       adsurl = {https://ui.adsabs.harvard.edu/abs/2024MNRAS.535..171P}
}

@ARTICLE{2026A&A...706A..30O,
       author = {{O'Donovan}, Barry and {Bitsch}, Bertram},
        title = "{Heavy-element-enriched atmospheres and where they are born}",
      journal = {\aap},
         year = 2026,
        month = jan,
       volume = {706},
          eid = {A30},
        pages = {A30},
          doi = {10.1051/0004-6361/202556898},
archivePrefix = {arXiv},
       eprint = {2512.07944},
 primaryClass = {astro-ph.EP},
       adsurl = {https://ui.adsabs.harvard.edu/abs/2026A&A...706A..30O}
}

@ARTICLE{2014AJ....148...54H,
       author = {{Hinkel}, Natalie R. and {Timmes}, F.~X. and {Young}, Patrick A. and {Pagano}, Michael D. and {Turnbull}, Margaret C.},
        title = "{Stellar Abundances in the Solar Neighborhood: The Hypatia Catalog}",
      journal = {\aj},
         year = 2014,
        month = sep,
       volume = {148},
       number = {3},
          eid = {54},
        pages = {54},
          doi = {10.1088/0004-6256/148/3/54},
archivePrefix = {arXiv},
       eprint = {1405.6719},
 primaryClass = {astro-ph.SR},
       adsurl = {https://ui.adsabs.harvard.edu/abs/2014AJ....148...54H}
}

@ARTICLE{2025A&A...699A.227H,
       author = {{Houge}, Adrien and {Johansen}, Anders and {Bergin}, Edwin and {Ciesla}, Fred J. and {Bitsch}, Bertram and {Lambrechts}, Michiel and {Henning}, Thomas and {Perotti}, Giulia},
        title = "{Burned to ashes: How the thermal decomposition of refractory organics in the inner protoplanetary disc impacts the gas-phase C/O ratio}",
      journal = {\aap},
         year = 2025,
        month = jul,
       volume = {699},
          eid = {A227},
        pages = {A227},
          doi = {10.1051/0004-6361/202555164},
archivePrefix = {arXiv},
       eprint = {2505.20427},
 primaryClass = {astro-ph.EP},
       adsurl = {https://ui.adsabs.harvard.edu/abs/2025A&A...699A.227H}
}

@ARTICLE{2025MNRAS.544.3562W,
       author = {{Williams}, Joe and {Krijt}, Sebastiaan and {Bitsch}, Bertram and {Houge}, Adrien and {Bergner}, Jennifer},
        title = "{Locked in ice: how pebble drift and volatile entrapment can significantly impact carbon and oxygen ratios in evolving protoplanetary discs}",
      journal = {\mnras},
         year = 2025,
        month = dec,
       volume = {544},
       number = {4},
        pages = {3562-3578},
          doi = {10.1093/mnras/staf1839},
archivePrefix = {arXiv},
       eprint = {2510.18587},
 primaryClass = {astro-ph.EP},
       adsurl = {https://ui.adsabs.harvard.edu/abs/2025MNRAS.544.3562W}
}

@ARTICLE{2025AandA...699A.342B,
       author = {{Bardet}, Deborah and {Changeat}, Quentin and {Venot}, Olivia and {Panek}, Emilie},
        title = "{Re-analysis of ten hot-Jupiter atmospheres with disequilibrium chemistry retrieval}",
      journal = {\aap},
         year = 2025,
        month = jul,
       volume = {699},
          eid = {A342},
        pages = {A342},
          doi = {10.1051/0004-6361/202453518},
archivePrefix = {arXiv},
       eprint = {2506.12806},
 primaryClass = {astro-ph.EP},
       adsurl = {https://ui.adsabs.harvard.edu/abs/2025A&A...699A.342B}
}

@ARTICLE{2012A&A...539A.148B,
       author = {{Birnstiel}, T. and {Klahr}, H. and {Ercolano}, B.},
        title = "{A simple model for the evolution of the dust population in protoplanetary disks}",
      journal = {\aap},
         year = 2012,
        month = mar,
       volume = {539},
          eid = {A148},
        pages = {A148},
          doi = {10.1051/0004-6361/201118136},
archivePrefix = {arXiv},
       eprint = {1201.5781},
 primaryClass = {astro-ph.EP},
       adsurl = {https://ui.adsabs.harvard.edu/abs/2012A&A...539A.148B}
}

@ARTICLE{2021MNRAS.501.2017N,
       author = {{Ndugu}, N. and {Bitsch}, B. and {Morbidelli}, A. and {Crida}, A. and {Jurua}, E.},
        title = "{Probing the impact of varied migration and gas accretion rates for the formation of giant planets in the pebble accretion scenario}",
      journal = {\mnras},
         year = 2021,
        month = feb,
       volume = {501},
       number = {2},
        pages = {2017-2028},
          doi = {10.1093/mnras/staa3629},
archivePrefix = {arXiv},
       eprint = {2011.09146},
 primaryClass = {astro-ph.EP},
       adsurl = {https://ui.adsabs.harvard.edu/abs/2021MNRAS.501.2017N}
}

@INPROCEEDINGS{2014prpl.conf..547J,
       author = {{Johansen}, A. and {Blum}, J. and {Tanaka}, H. and {Ormel}, C. and {Bizzarro}, M. and {Rickman}, H.},
        title = "{The Multifaceted Planetesimal Formation Process}",
    booktitle = {Protostars and Planets VI},
         year = 2014,
       editor = {{Beuther}, Henrik and {Klessen}, Ralf S. and {Dullemond}, Cornelis P. and {Henning}, Thomas},
        month = jan,
        pages = {547-570},
          doi = {10.2458/azu_uapress_9780816531240-ch024},
archivePrefix = {arXiv},
       eprint = {1402.1344},
 primaryClass = {astro-ph.EP},
       adsurl = {https://ui.adsabs.harvard.edu/abs/2014prpl.conf..547J}
}

@ARTICLE{2011MNRAS.410..293P,
       author = {{Paardekooper}, S.-J. and {Baruteau}, C. and {Kley}, W.},
        title = "{A torque formula for non-isothermal Type I planetary migration - II. Effects of diffusion}",
      journal = {\mnras},
         year = 2011,
        month = jan,
       volume = {410},
       number = {1},
        pages = {293-303},
          doi = {10.1111/j.1365-2966.2010.17442.x},
archivePrefix = {arXiv},
       eprint = {1007.4964},
 primaryClass = {astro-ph.EP},
       adsurl = {https://ui.adsabs.harvard.edu/abs/2011MNRAS.410..293P}
}

@ARTICLE{2021Natur.598..580L,
       author = {{Line}, Michael R. and {Brogi}, Matteo and {Bean}, Jacob L. and {Gandhi}, Siddharth and {Zalesky}, Joseph and {Parmentier}, Vivien and {Smith}, Peter and {Mace}, Gregory N. and {Mansfield}, Megan and {Kempton}, Eliza M.-R. and {Fortney}, Jonathan J. and {Shkolnik}, Evgenya and {Patience}, Jennifer and {Rauscher}, Emily and {D{\'e}sert}, Jean-Michel and {Wardenier}, Joost P.},
        title = "{A solar C/O and sub-solar metallicity in a hot Jupiter atmosphere}",
      journal = {\nat},
         year = 2021,
        month = oct,
       volume = {598},
       number = {7882},
        pages = {580-584},
          doi = {10.1038/s41586-021-03912-6},
archivePrefix = {arXiv},
       eprint = {2110.14821},
 primaryClass = {astro-ph.EP},
       adsurl = {https://ui.adsabs.harvard.edu/abs/2021Natur.598..580L}
}

@ARTICLE{2025NatAs...9..845E,
       author = {{Evans-Soma}, Thomas M. and {Sing}, David K. and {Barstow}, Joanna K. and {Piette}, Anjali A.~A. and {Taylor}, Jake and {Lothringer}, Joshua D. and {Reggiani}, Henrique and {Goyal}, Jayesh M. and {Ahrer}, Eva-Maria and {Mayne}, Nathan J. and {Rustamkulov}, Zafar and {Kataria}, Tiffany and {Christie}, Duncan A. and {Gapp}, Cyril and {Dong}, Jiayin and {Foreman-Mackey}, Daniel and {Hattori}, Soichiro and {Marley}, Mark S.},
        title = "{SiO and a super-stellar C/O ratio in the atmosphere of the giant exoplanet WASP-121 b}",
      journal = {Nature Astronomy},
         year = 2025,
        month = jun,
       volume = {9},
        pages = {845-861},
          doi = {10.1038/s41550-025-02513-x},
archivePrefix = {arXiv},
       eprint = {2506.01771},
 primaryClass = {astro-ph.EP},
       adsurl = {https://ui.adsabs.harvard.edu/abs/2025NatAs...9..845E}
}

@INPROCEEDINGS{2021atat.confE..20P,
       author = {{Pelletier}, Stefan},
        title = "{Where is the Water? Jupiter-like C/H ratio but strong H2O depletion found on tau Boo b using SPIRou}",
    booktitle = {Atmospheres, Atmospheres! Do I Look Like I Care About Atmospheres?},
         year = 2021,
        month = oct,
          eid = {20},
        pages = {20},
          doi = {10.5281/zenodo.5548123},
       adsurl = {https://ui.adsabs.harvard.edu/abs/2021atat.confE..20P}
}

@ARTICLE{2026A&A...707A.276G,
       author = {{Guzm{\'a}n Franco}, Angie Daniela and {Savvidou}, Sofia and {Bitsch}, Bertram},
        title = "{How initial disc conditions sculpt the atmospheric composition of giant planets}",
      journal = {\aap},
         year = 2026,
        month = mar,
       volume = {707},
          eid = {A276},
        pages = {A276},
          doi = {10.1051/0004-6361/202556632},
archivePrefix = {arXiv},
       eprint = {2601.10285},
 primaryClass = {astro-ph.EP},
       adsurl = {https://ui.adsabs.harvard.edu/abs/2026A&A...707A.276G}
}

@ARTICLE{2024A&A...691A..50E,
       author = {{Eberlein}, Mark and {Bitsch}, Bertram and {Helled}, Ravit},
        title = "{Disc and atmosphere composition of multi-planet systems}",
      journal = {\aap},
         year = 2024,
        month = nov,
       volume = {691},
          eid = {A50},
        pages = {A50},
          doi = {10.1051/0004-6361/202449840},
archivePrefix = {arXiv},
       eprint = {2407.20117},
 primaryClass = {astro-ph.EP},
       adsurl = {https://ui.adsabs.harvard.edu/abs/2024A&A...691A..50E}
}

@INPROCEEDINGS{2009AIPC.1158....3M,
       author = {{Mamajek}, Eric E.},
        title = "{Initial Conditions of Planet Formation: Lifetimes of Primordial Disks}",
    booktitle = {Exoplanets and Disks: Their Formation and Diversity},
         year = 2009,
       editor = {{Usuda}, Tomonori and {Tamura}, Motohide and {Ishii}, Miki},
       series = {American Institute of Physics Conference Series},
       volume = {1158},
        month = aug,
    publisher = {AIP},
        pages = {3-10},
          doi = {10.1063/1.3215910},
archivePrefix = {arXiv},
       eprint = {0906.5011},
 primaryClass = {astro-ph.EP},
       adsurl = {https://ui.adsabs.harvard.edu/abs/2009AIPC.1158....3M}
}

@ARTICLE{2022ApJ...939L..10P,
       author = {{Pfalzner}, Susanne and {Dehghani}, Shahrzad and {Michel}, Arnaud},
        title = "{Most Planets Might Have More than 5 Myr of Time to Form}",
      journal = {\apjl},
         year = 2022,
        month = nov,
       volume = {939},
       number = {1},
          eid = {L10},
        pages = {L10},
          doi = {10.3847/2041-8213/ac9839},
archivePrefix = {arXiv},
       eprint = {2210.02420},
 primaryClass = {astro-ph.GA},
       adsurl = {https://ui.adsabs.harvard.edu/abs/2022ApJ...939L..10P}
}

@ARTICLE{2013MNRAS.431.3494L,
       author = {{Lega}, E. and {Morbidelli}, A. and {Nesvorn{\'y}}, D.},
        title = "{Early dynamical instabilities in the giant planet systems}",
      journal = {\mnras},
         year = 2013,
        month = jun,
       volume = {431},
       number = {4},
        pages = {3494-3500},
          doi = {10.1093/mnras/stt431},
archivePrefix = {arXiv},
       eprint = {1303.6062},
 primaryClass = {astro-ph.EP},
       adsurl = {https://ui.adsabs.harvard.edu/abs/2013MNRAS.431.3494L}
}

@ARTICLE{2023A&A...674A.178B,
       author = {{Bitsch}, Bertram and {Izidoro}, Andre},
        title = "{Giants are bullies: How their growth influences systems of inner sub-Neptunes and super-Earths}",
      journal = {\aap},
         year = 2023,
        month = jun,
       volume = {674},
          eid = {A178},
        pages = {A178},
          doi = {10.1051/0004-6361/202245040},
archivePrefix = {arXiv},
       eprint = {2304.12758},
 primaryClass = {astro-ph.EP},
       adsurl = {https://ui.adsabs.harvard.edu/abs/2023A&A...674A.178B}
}

@ARTICLE{2012Natur.491..418B,
       author = {{Batygin}, Konstantin},
        title = "{A primordial origin for misalignments between stellar spin axes and planetary orbits}",
      journal = {\nat},
         year = 2012,
        month = nov,
       volume = {491},
       number = {7424},
        pages = {418-420},
          doi = {10.1038/nature11560},
       adsurl = {https://ui.adsabs.harvard.edu/abs/2012Natur.491..418B}
}

@ARTICLE{2008ApJ...678.1396J,
       author = {{Jackson}, Brian and {Greenberg}, Richard and {Barnes}, Rory},
        title = "{Tidal Evolution of Close-in Extrasolar Planets}",
      journal = {\apj},
         year = 2008,
        month = may,
       volume = {678},
       number = {2},
        pages = {1396-1406},
          doi = {10.1086/529187},
archivePrefix = {arXiv},
       eprint = {0802.1543},
 primaryClass = {astro-ph},
       adsurl = {https://ui.adsabs.harvard.edu/abs/2008ApJ...678.1396J}
}

@ARTICLE{2012MNRAS.423..486L,
       author = {{Lai}, Dong},
        title = "{Tidal dissipation in planet-hosting stars: damping of spin-orbit misalignment and survival of hot Jupiters}",
      journal = {\mnras},
         year = 2012,
        month = jun,
       volume = {423},
       number = {1},
        pages = {486-492},
          doi = {10.1111/j.1365-2966.2012.20893.x},
archivePrefix = {arXiv},
       eprint = {1109.4703},
 primaryClass = {astro-ph.EP},
       adsurl = {https://ui.adsabs.harvard.edu/abs/2012MNRAS.423..486L}
}

@ARTICLE{2010ApJ...718L.145W,
       author = {{Winn}, Joshua N. and {Fabrycky}, Daniel and {Albrecht}, Simon and {Johnson}, John Asher},
        title = "{Hot Stars with Hot Jupiters Have High Obliquities}",
      journal = {\apjl},
         year = 2010,
        month = aug,
       volume = {718},
       number = {2},
        pages = {L145-L149},
          doi = {10.1088/2041-8205/718/2/L145},
archivePrefix = {arXiv},
       eprint = {1006.4161},
 primaryClass = {astro-ph.EP},
       adsurl = {https://ui.adsabs.harvard.edu/abs/2010ApJ...718L.145W}
}

@ARTICLE{2001A&A...366..263P,
       author = {{Papaloizou}, J.~C.~B. and {Nelson}, R.~P. and {Masset}, F.},
        title = "{Orbital eccentricity growth through disc-companion tidal interaction}",
      journal = {\aap},
         year = 2001,
        month = jan,
       volume = {366},
        pages = {263-275},
          doi = {10.1051/0004-6361:20000011},
       adsurl = {https://ui.adsabs.harvard.edu/abs/2001A&A...366..263P}
}

@ARTICLE{2006A&A...447..369K,
       author = {{Kley}, W. and {Dirksen}, G.},
        title = "{Disk eccentricity and embedded planets}",
      journal = {\aap},
         year = 2006,
        month = feb,
       volume = {447},
       number = {1},
        pages = {369-377},
          doi = {10.1051/0004-6361:20053914},
archivePrefix = {arXiv},
       eprint = {astro-ph/0510393},
 primaryClass = {astro-ph},
       adsurl = {https://ui.adsabs.harvard.edu/abs/2006A&A...447..369K}
}

@ARTICLE{2013A&A...555A.124B,
       author = {{Bitsch}, B. and {Crida}, A. and {Libert}, A.-S. and {Lega}, E.},
        title = "{Highly inclined and eccentric massive planets. I. Planet-disc interactions}",
      journal = {\aap},
         year = 2013,
        month = jul,
       volume = {555},
          eid = {A124},
        pages = {A124},
          doi = {10.1051/0004-6361/201220310},
archivePrefix = {arXiv},
       eprint = {1305.7330},
 primaryClass = {astro-ph.EP},
       adsurl = {https://ui.adsabs.harvard.edu/abs/2013A&A...555A.124B}
}

@ARTICLE{2008ApJ...686..603J,
       author = {{Juri{\'c}}, Mario and {Tremaine}, Scott},
        title = "{Dynamical Origin of Extrasolar Planet Eccentricity Distribution}",
      journal = {\apj},
         year = 2008,
        month = oct,
       volume = {686},
       number = {1},
        pages = {603-620},
          doi = {10.1086/590047},
archivePrefix = {arXiv},
       eprint = {astro-ph/0703160},
 primaryClass = {astro-ph},
       adsurl = {https://ui.adsabs.harvard.edu/abs/2008ApJ...686..603J}
}

@ARTICLE{2011MNRAS.417.2166S,
       author = {{Southworth}, John},
        title = "{Homogeneous studies of transiting extrasolar planets - IV. Thirty systems with space-based light curves}",
      journal = {\mnras},
         year = 2011,
        month = nov,
       volume = {417},
       number = {3},
        pages = {2166-2196},
          doi = {10.1111/j.1365-2966.2011.19399.x},
archivePrefix = {arXiv},
       eprint = {1107.1235},
 primaryClass = {astro-ph.EP},
       adsurl = {https://ui.adsabs.harvard.edu/abs/2011MNRAS.417.2166S}
}

@ARTICLE{2018ApJ...856...37B,
       author = {{Buchhave}, Lars A. and {Bitsch}, Bertram and {Johansen}, Anders and {Latham}, David W. and {Bizzarro}, Martin and {Bieryla}, Allyson and {Kipping}, David M.},
        title = "{Jupiter Analogs Orbit Stars with an Average Metallicity Close to That of the Sun}",
      journal = {\apj},
         year = 2018,
        month = mar,
       volume = {856},
       number = {1},
          eid = {37},
        pages = {37},
          doi = {10.3847/1538-4357/aaafca},
archivePrefix = {arXiv},
       eprint = {1802.06794},
 primaryClass = {astro-ph.EP},
       adsurl = {https://ui.adsabs.harvard.edu/abs/2018ApJ...856...37B}
}

@ARTICLE{2018ApJ...864...78K,
       author = {{Krijt}, Sebastiaan and {Schwarz}, Kamber R. and {Bergin}, Edwin A. and {Ciesla}, Fred J.},
        title = "{Transport of CO in Protoplanetary Disks: Consequences of Pebble Formation, Settling, and Radial Drift}",
      journal = {\apj},
         year = 2018,
        month = sep,
       volume = {864},
       number = {1},
          eid = {78},
        pages = {78},
          doi = {10.3847/1538-4357/aad69b},
archivePrefix = {arXiv},
       eprint = {1808.01840},
 primaryClass = {astro-ph.EP},
       adsurl = {https://ui.adsabs.harvard.edu/abs/2018ApJ...864...78K}
}

@ARTICLE{2024ApJ...975..166B,
       author = {{Bergner}, Jennifer B. and {Sturm}, J.~A. and {Piacentino}, Elettra L. and {McClure}, M.~K. and {{\"O}berg}, Karin I. and {Boogert}, A.~C.~A. and {Dartois}, E. and {Drozdovskaya}, M.~N. and {Fraser}, H.~J. and {Harsono}, Daniel and {Ioppolo}, Sergio and {Law}, Charles J. and {Lis}, Dariusz C. and {McGuire}, Brett A. and {Melnick}, Gary J. and {Noble}, Jennifer A. and {Palumbo}, M.~E. and {Pendleton}, Yvonne J. and {Perotti}, Giulia and {Qasim}, Danna and {Rocha}, W.~R.~M. and {van Dishoeck}, E.~F.},
        title = "{JWST Ice Band Profiles Reveal Mixed Ice Compositions in the HH 48 NE Disk}",
      journal = {\apj},
         year = 2024,
        month = nov,
       volume = {975},
       number = {2},
          eid = {166},
        pages = {166},
          doi = {10.3847/1538-4357/ad79fc},
archivePrefix = {arXiv},
       eprint = {2409.08117},
 primaryClass = {astro-ph.EP},
       adsurl = {https://ui.adsabs.harvard.edu/abs/2024ApJ...975..166B}
}

@ARTICLE{2025AJ....170..299K,
       author = {{Kawai}, Yugo and {Fukui}, Akihiko and {Watanabe}, Noriharu and {Fukazawa}, Sho and {Narita}, Norio},
        title = "{Identifying Close-in Jupiters that Arrived via Disk Migration: Evidence of Primordial Alignment, Preference of Nearby Companions and Hint of Runaway Migration}",
      journal = {\aj},
         year = 2025,
        month = dec,
       volume = {170},
       number = {6},
          eid = {299},
        pages = {299},
          doi = {10.3847/1538-3881/ae0a11},
archivePrefix = {arXiv},
       eprint = {2509.16322},
 primaryClass = {astro-ph.EP},
       adsurl = {https://ui.adsabs.harvard.edu/abs/2025AJ....170..299K}
}

@ARTICLE{2023A&A...677L...7M,
       author = {{Mah}, Jingyi and {Bitsch}, Bertram and {Pascucci}, Ilaria and {Henning}, Thomas},
        title = "{Close-in ice lines and the super-stellar C/O ratio in discs around very low-mass stars}",
      journal = {\aap},
         year = 2023,
        month = sep,
       volume = {677},
          eid = {L7},
        pages = {L7},
          doi = {10.1051/0004-6361/202347169},
archivePrefix = {arXiv},
       eprint = {2308.15128},
 primaryClass = {astro-ph.EP},
       adsurl = {https://ui.adsabs.harvard.edu/abs/2023A&A...677L...7M}
}

@ARTICLE{2012MNRAS.419..366M,
       author = {{Moeckel}, Nickolas and {Armitage}, Philip J.},
        title = "{Hydrodynamic outcomes of planet scattering in transitional discs}",
      journal = {\mnras},
         year = 2012,
        month = jan,
       volume = {419},
       number = {1},
        pages = {366-376},
          doi = {10.1111/j.1365-2966.2011.19699.x},
archivePrefix = {arXiv},
       eprint = {1108.5382},
 primaryClass = {astro-ph.EP},
       adsurl = {https://ui.adsabs.harvard.edu/abs/2012MNRAS.419..366M}
}

@ARTICLE{2013ApJ...767L..24D,
       author = {{Dawson}, Rebekah I. and {Murray-Clay}, Ruth A.},
        title = "{Giant Planets Orbiting Metal-rich Stars Show Signatures of Planet-Planet Interactions}",
      journal = {\apjl},
         year = 2013,
        month = apr,
       volume = {767},
       number = {2},
          eid = {L24},
        pages = {L24},
          doi = {10.1088/2041-8205/767/2/L24},
archivePrefix = {arXiv},
       eprint = {1302.6244},
 primaryClass = {astro-ph.EP},
       adsurl = {https://ui.adsabs.harvard.edu/abs/2013ApJ...767L..24D}
}

\begin{appendix}
\nolinenumbers

\onecolumn
\section{Chemical evolution of the disc}\label{appendixA}

\begin{figure}[ht!]
\centering
\includegraphics[width=\columnwidth]{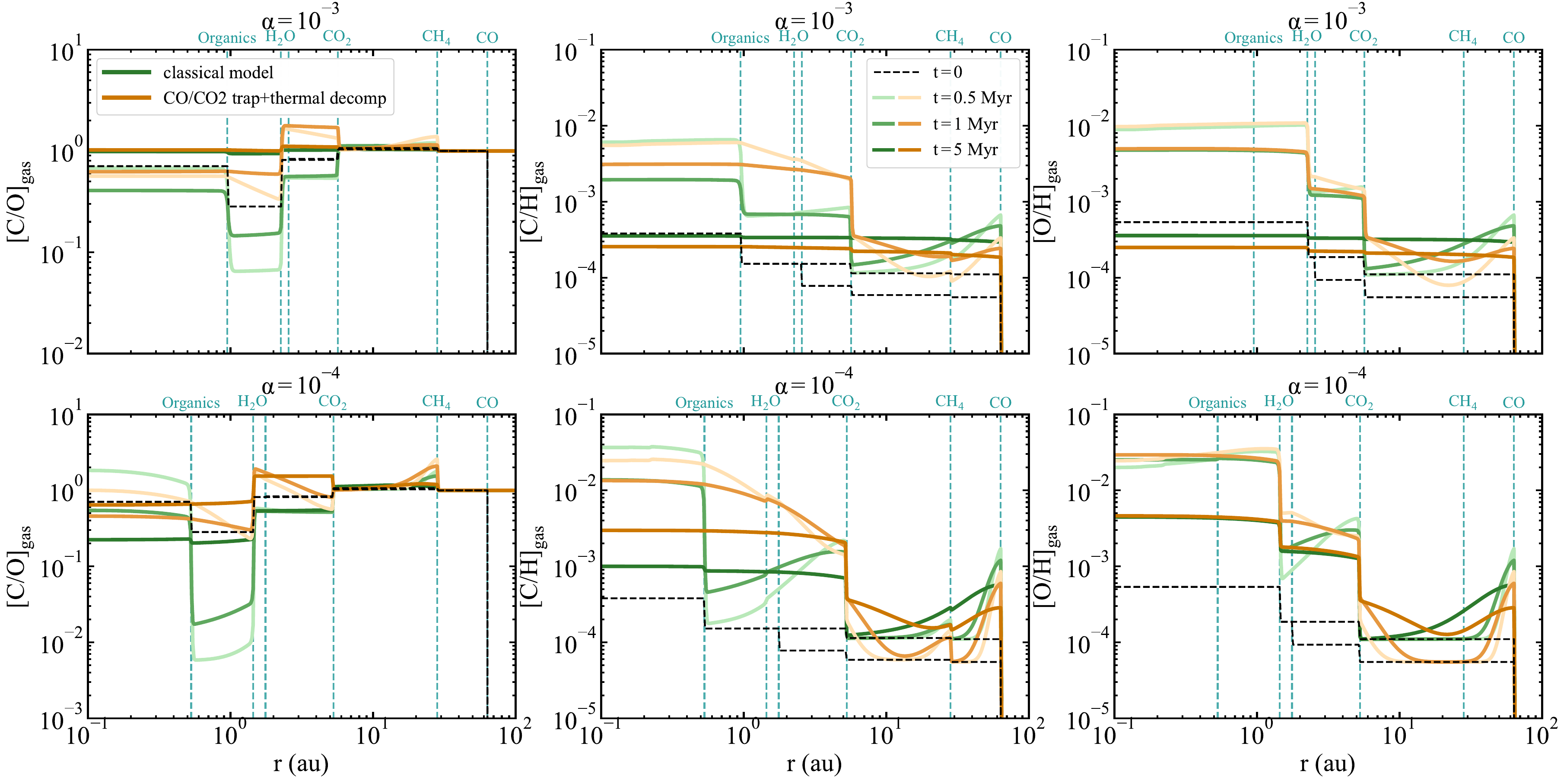}
\caption{Radial chemical composition of the gas in a representative solar-composition protoplanetary disc without embedded planets for two different viscosity parameters ($\alpha = 10^{-3}$ and $\alpha = 10^{-4}$).
This figure is intended to illustrate the effect of the additional chemical processes
included in our model. The planet-formation simulations for individual
systems instead adopted the corresponding host-star abundances listed in
Table~\ref{tab:tableC1}. The three panels show the radial
distributions of C/O, C/H, and O/H in the gas phase. The green curves
correspond to the classical chemical model, while the orange curves include
the additional processes of CO/CO$_2$ trapping in water ice and the
thermal decomposition of refractory organics. Different line shades
indicate different evolutionary times of the disc ($t=0$, $0.5$, $1$,
and $5\,\mathrm{Myr}$). Vertical dashed lines mark the locations of the
major condensation fronts of the main volatile species. The vertical line just exterior to the water evaporation front corresponds to the ``volcano line'', where CO and CO$_2$ are released from water ice.}
\label{fig:figA1}
\end{figure}

\begin{figure}[ht!]
\centering
\includegraphics[width=0.6\columnwidth, height=0.6\textheight, keepaspectratio]{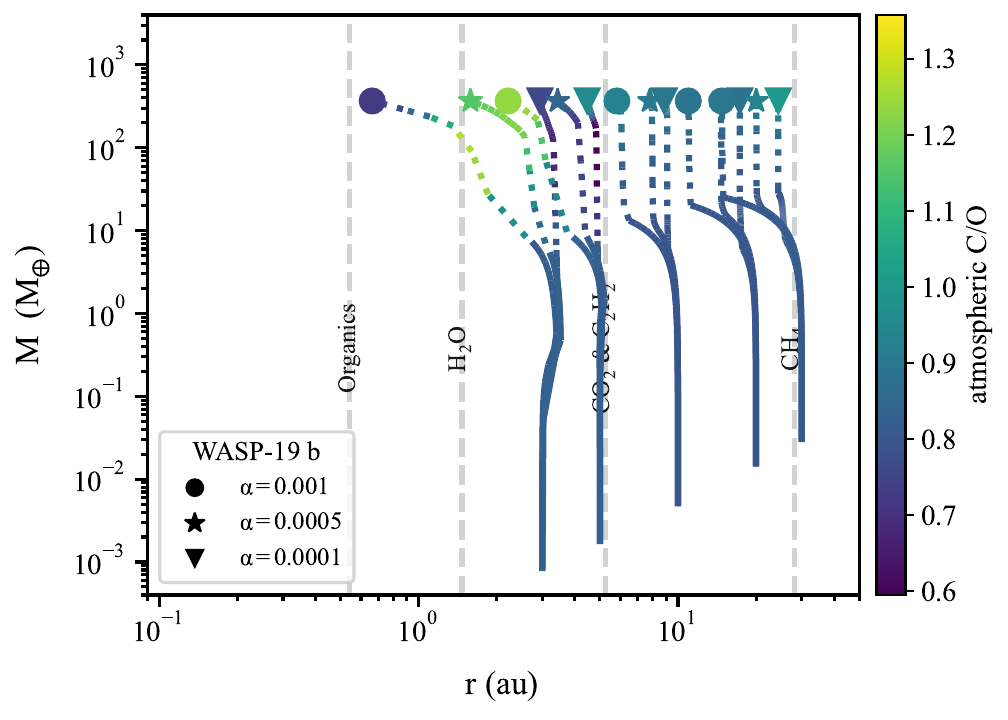}
\caption{Planet growth and migration tracks for the formation of WASP-19~b starting from different initial orbital locations. Solid parts of the tracks indicate the pebble-accretion phase before the planet reached pebble isolation mass, while dotted parts indicate the subsequent gas-accretion phase. Different symbols indicate different disc viscosities ($\alpha = 10^{-3}$, $5\times10^{-4}$, and $10^{-4}$). The colour of each track represents the resulting atmospheric C/O ratio of the planet.}
\label{fig:figA2}
\end{figure}

Figure~\ref{fig:figA1} shows the radial distribution of gas-phase chemical abundances in a representative solar-composition disc. The purpose of this figure is to illustrate how the additional chemical processes included in our model modify the radial gas-phase C/O, C/H, and O/H profiles. The reference disc is a viscously evolving disc around a solar-type star ($M_\star = 1\,M_\odot$), with a disc mass $M_0 = 0.128\,M_\odot$, a disc radius $R_0 = 137$ au, and an initial dust-to-gas ratio of 0.02. The initial chemical composition followed a standard carbon partitioning, where 60\% of carbon is locked in refractory material, while the remaining carbon is distributed among volatile species (29\% in CO, 10\% in CO$_2$, and 1\% in CH$_4$).

We compared the classical chemical model with a model that included CO/CO$_2$ trapping in amorphous water ice, adopting a fiducial trapping efficiency of 50\% for both CO and CO$_2$ \citep{2025MNRAS.544.3562W}, and the thermal decomposition of refractory organics \citep{2025A&A...699A.227H}.
These additional processes significantly modified the radial distribution of carbon and oxygen, particularly near the major condensation fronts. In the planet-formation simulations used to interpret the observed hot Jupiters, the same chemical framework was
applied to discs whose elemental abundances and dust-to-gas ratios were scaled to the corresponding host-star abundances, as described in Section~\ref{model} and listed in Table~\ref{tab:tableC1}.

Figure~\ref{fig:figA2} shows the evolution of planetary mass as a function of orbital separation during the formation process. 
The figure illustrates how the atmospheric composition of a giant planet is shaped by both its formation location and its migration history. During the pebble-accretion phase, the C/O ratio was set by the composition of the accreted solids; after reaching the pebble isolation mass, planets entered the gas-accretion phase, during which the atmospheric C/O ratio was affected by the composition of the accreted gas. Planets that crossed condensation fronts during their growth experienced transitions in the composition of the accreted gas, leading to changes in their atmospheric C/O ratios. 
For example, a planet forming beyond the H$_2$O snowline initially accreted relatively oxygen-poor gas, also enriched by outward-diffusing C$_2$H$_2$, and thus attained a high C/O ratio. As it migrated inward and crossed the H$_2$O snowline, the evaporation of water ice enriched the gas phase in oxygen, resulting in a decrease in the planetary C/O ratio. Planets that accreted gas in different regions, or that experienced different migration pathways, developed distinct atmospheric compositions.

\section{Atmospheric composition constraints}\label{appendixB}

\begin{figure}[ht!]
\centering
\includegraphics[width=\columnwidth]{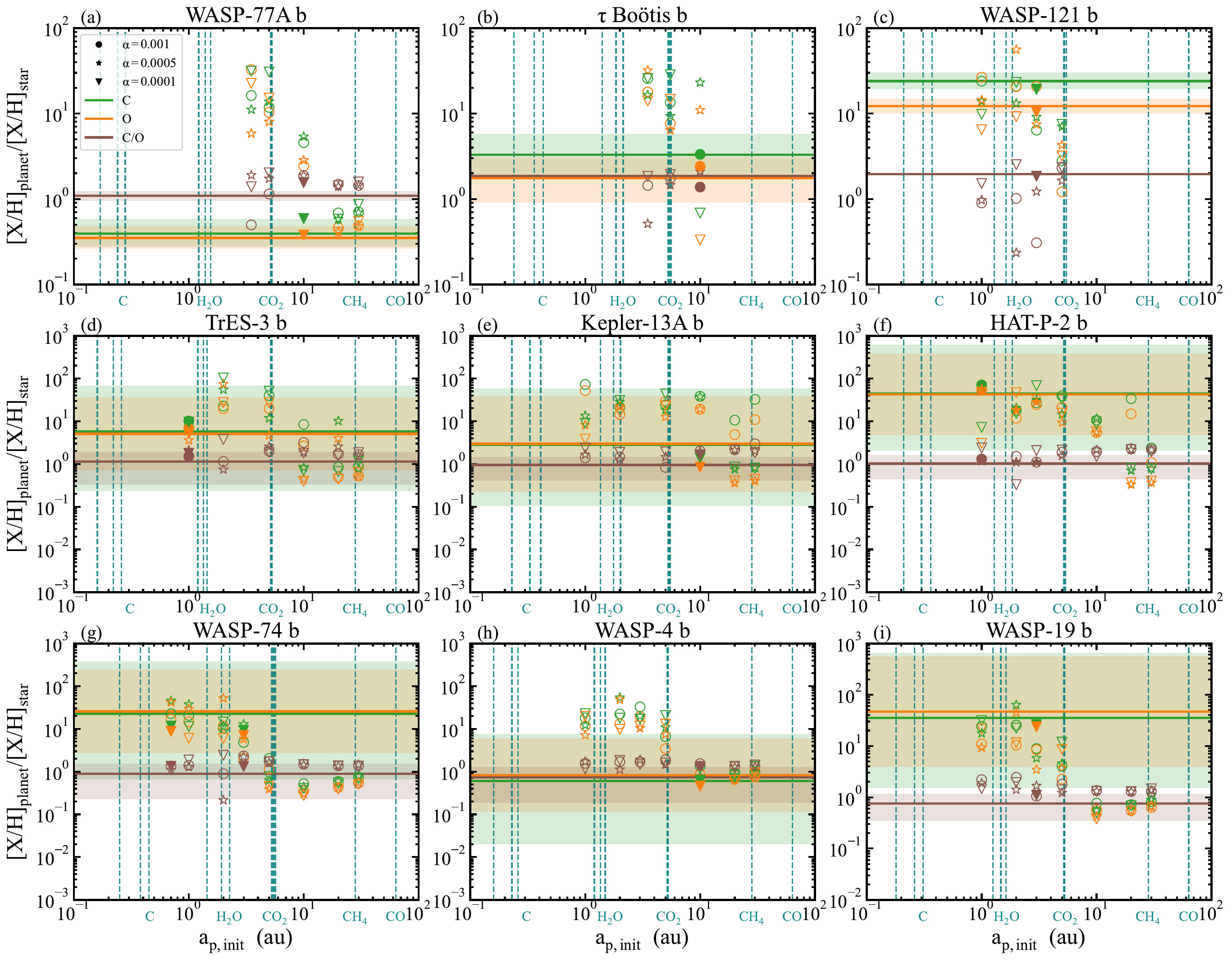}
\caption{Simulated atmospheric abundances for the nine hot Jupiter systems considered in this work. Each panel shows the elemental abundances (C/H, O/H, and C/O) of planets formed from different initial orbital locations in the disc. The symbols represent the results of individual simulations with different disc viscosities. Open symbols show all simulations, while filled symbols highlight the simulations that simultaneously reproduced the observed C/H and O/H abundances and reached the observed planetary mass at $t \geq 1\,\mathrm{Myr}$. These filled symbols are the simulations used to infer the formation regions shown in Fig.~\ref{fig:fig1}. The horizontal shaded regions indicate the observational constraints for each system. Vertical dashed lines mark the locations of the major snowlines in the protoplanetary disc. The three lines for each snowline correspond to different disc viscosities, which give different disc temperatures and different snowline locations.}
\label{fig:figB1}
\end{figure}

\begin{figure}[ht!]
\centering
\includegraphics[width=0.6\columnwidth, height=0.6\textheight, keepaspectratio]{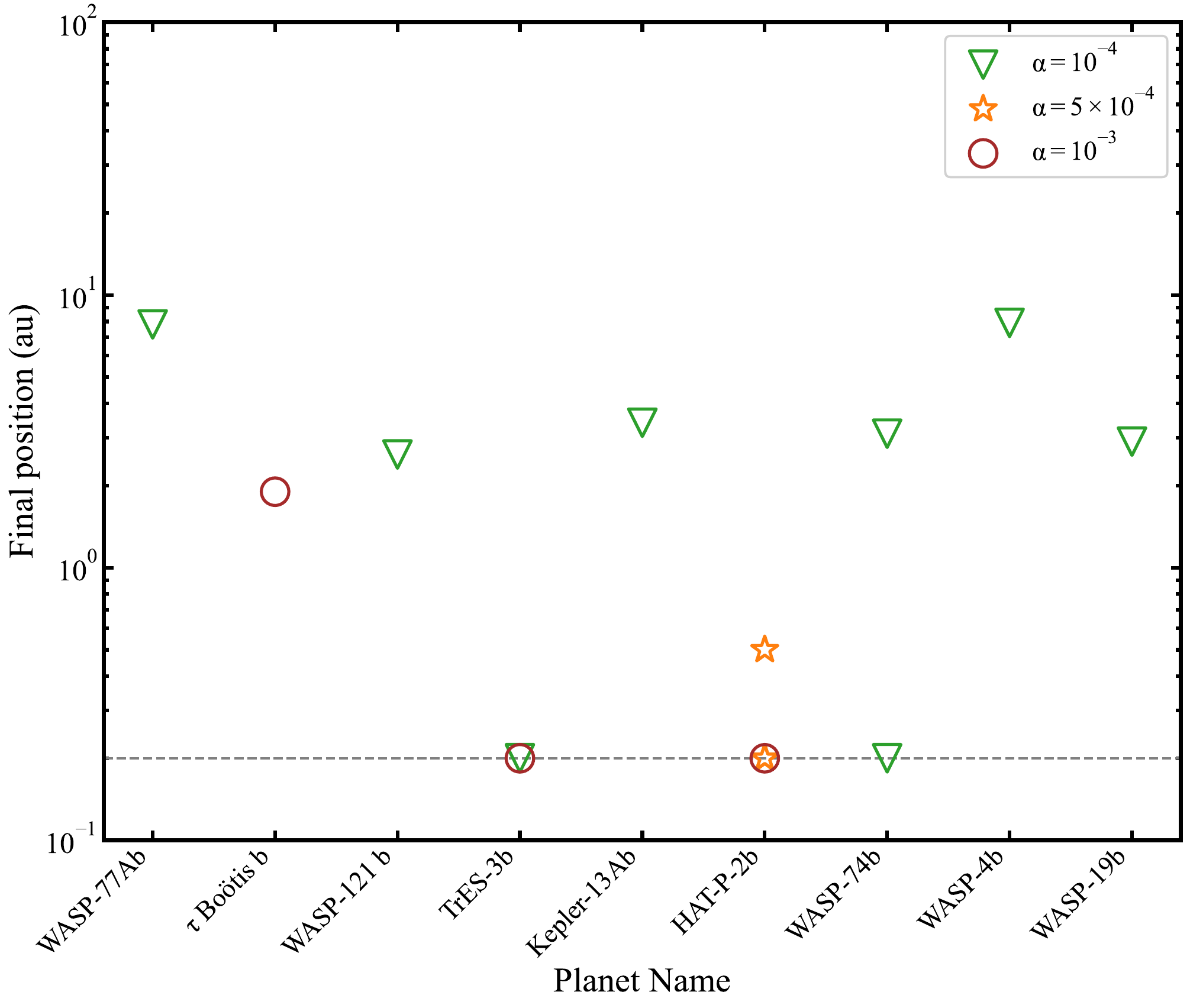}
\caption{Final orbital positions of the selected simulations that reproduced the observed atmospheric abundances of the nine hot Jupiter systems and reached the observed planetary masses at $t \geq 1\,\mathrm{Myr}$. The horizontal dashed line marks the inner boundary of simulations at $0.2\,\mathrm{au}$.}
\label{fig:figB2}
\end{figure}

Figure~\ref{fig:figB1} shows the atmospheric abundances predicted by our planet formation simulations for the nine hot Jupiter systems considered in this work. For each system, we performed simulations starting from a range of initial orbital locations and computed the resulting atmospheric elemental abundances (C/H, O/H, and C/O) of the planets.

Each point represents the outcome of an individual simulation with a specific initial orbital location and disc viscosity. Open symbols show all simulation results, whereas filled symbols indicate the subset used for the formation-location inference. These selected simulations reproduced the observed C/H and O/H abundances and reached the observed planetary mass at $t \geq 1\,\mathrm{Myr}$. This time criterion excluded results in which planets grew to their final masses too early compared with typical protoplanetary disc lifetimes \citep{2009AIPC.1158....3M,2022ApJ...939L..10P}. Some high-viscosity points can match the atmospheric abundances but were not included in the final formation-location classification. This mainly affects lower-mass planets, for which high-viscosity discs can lead to rapid growth before $1\,\mathrm{Myr}$. In other systems, such as HAT-P-2~b, TrES-3~b, and $\tau$~Boötis~b, high-viscosity simulations are instead needed to reach the observed planetary masses within a reasonable time, because gas accretion, which is limited by the viscous disc accretion rate, can progress faster.

Figure~\ref{fig:figB1} highlights a limitation of using the atmospheric C/O ratio alone as a diagnostic of planet formation. In our model, the atmospheric composition depends on both the initial orbital location and the migration history of the planet, with the latter primarily controlled by the disc viscosity. As a result, multiple combinations of initial position and migration efficiency can produce similar C/O ratios, making it difficult to uniquely constrain the formation location using C/O alone. In contrast, the combined constraints from C/H and O/H significantly reduce this degeneracy and allow for a more robust inference of the planet’s origin.

We show in Fig.~\ref{fig:figB2} the final positions of the selected simulations that reproduced the observed atmospheric abundances and satisfied the formation-time criterion. Planets that reached the inner disc boundary can be interpreted as systems for which disc-driven migration is sufficient to bring the planet close to its present-day orbit. In contrast, planets that remained outside this boundary at the end of the simulation require additional dynamical evolution, such as planet--planet scattering, to reach their observed orbital separations. These results provide a direct link between the atmospheric composition constraints and the inferred migration pathways of hot Jupiters. The implications for the orbital obliquities and eccentricities of these planets are discussed in Table~\ref{tab:hot_jupiters} and Section~\ref{sec:discussion}.

\section{Model parameters}\label{appendixC}

We summarise in Table~\ref{tab:tableC1}(a) the stellar abundances and corresponding disc parameters adopted as input in the simulations, and in Table~\ref{tab:tableC1}(b) the observed atmospheric abundances used for comparison. 
For all systems, we adopted $M_{\rm disc}=0.07\,M_\star$. The fragmentation velocity was set to $v_{\rm frag}=5\,{\rm m\,s^{-1}}$, and the disc radius was computed as $R_{\rm disc}=30\,{\rm au}\,(M_\star/(0.1\,M_\odot))^{0.7}$.

\begin{table*}[t]
\caption{Input stellar and disc parameters, and observed planetary properties of the nine hot Jupiter systems.}
\label{tab:tableC1}
\centering
\footnotesize
\renewcommand{\arraystretch}{1.25}

\setlength{\tabcolsep}{4pt}
\begin{tabular}{lccccccccc}
\hline\hline
\multicolumn{10}{c}{(a) Host-star parameters and corresponding disc properties} \\
\hline
Name & $M_{\star}$  & {[C/H]} & {[O/H]} & {[Mg/H]} & {[Si/H]} & {[Fe/H]} & {[S/H]} & Dust-to-gas ratio & $R_{\rm disc}$  \\
 & ($M_\odot$) & & & & & & & &(au) \\
\hline
WASP-77A  & 1.00 & -0.01 & -0.01 & -0.07 & 0.00 & 0.00 & -0.24 & 0.016 & 150 \\
$\tau$~Boötis  & 1.38 & 0.29 & 0.30 & 0.16 & 0.34 & 0.26 & 0.24 & 0.030 & 188 \\
WASP-121  & 1.36 & 0.067 & 0.165 & 0.046 & 0.195 & 0.164 & 0.144 & 0.022 & 186 \\
TrES-3  & 0.928 & -0.17 & -0.034 & -0.15 & -0.14 & -0.23 & -0.12 & 0.013 & 143 \\
Kepler-13A  & 1.72 & 0.16 & 0.17 & 0.24 & 0.22 & 0.16 & 0.24 & 0.024 & 220 \\
HAT-P-2  & 1.33 & 0.105 & 0.14 & 0.174 & 0.16 & 0.10 & 0.18 & 0.022 & 184 \\
WASP-74  & 1.191 & 0.36 & 0.37 & 0.27 & 0.40 & 0.38 & 0.40 & 0.036 & 170 \\
WASP-4  & 0.89 & -0.03 & -0.10 & 0.00 & 0.02 & 0.00 & 0.02 & 0.014 & 139 \\
WASP-19  & 0.965 & 0.09 & -0.01 & 0.20 & 0.26 & 0.20 & 0.08 & 0.019 & 147 \\
\hline
\end{tabular}

\vspace{0.1cm}

\begin{tabular}{lccccc}
\hline\hline
\multicolumn{6}{c}{(b) Observed planetary properties} \\
\hline
Name &
$M_{\rm p}$  &
C/H  &
O/H &
C/O &
Reference \\
 & ($M_{\rm Jup}$) & ($\times10^{-4}$)  & ($\times10^{-4}$) & & \\

\hline

WASP-77A~b &
1.76 &
$1.03^{+0.50}_{-0.30}$ &
$1.72^{+0.64}_{-0.43}$ &
$0.59\pm0.08$ &
\citet{2021Natur.598..580L} \\

$\tau$~Boötis~b &
6.24 &
$17.3^{+13}_{-8.3}$ &
$17.2^{+13}_{-8.4}$ &
$1.00^{+0.01}_{-0.00}$ &
\citet{2021atat.confE..20P} \\

WASP-121~b &
1.20 &
$81^{+21}_{-16}$ &
$87^{+20}_{-16}$ &
$0.92^{+0.02}_{-0.03}$ &
\citet{2025NatAs...9..845E} \\

TrES-3~b &
1.801 &
$10.3^{+110}_{-9.9}$ &
$22.9^{+140}_{-20}$ &
$0.45\pm0.32$ &
\citet{2025AandA...699A.342B} \\

Kepler-13A~b &
9.28 &
$10.9^{+210}_{-10}$ &
$21.4^{+260}_{-20}$ &
$0.51\pm0.29$ &
\citet{2025AandA...699A.342B} \\

HAT-P-2~b &
9.02 &
$150^{+2000}_{-140}$ &
$288^{+2300}_{-260}$ &
$0.52\pm0.30$ &
\citet{2025AandA...699A.342B} \\

WASP-74~b &
0.826 &
$139^{+2200}_{-140}$ &
$295^{+2500}_{-260}$ &
$0.47\pm0.35$ &
\citet{2025AandA...699A.342B} \\

WASP-4~b &
1.17 &
$1.5^{+18}_{-1.5}$ &
$3.2^{+20}_{-2.8}$ &
$0.47\pm0.35$ &
\citet{2025AandA...699A.342B} \\

WASP-19~b &
1.154 &
$116^{+2000}_{-110}$ &
$224^{+2500}_{-210}$ &
$0.52\pm0.28$ &
\citet{2025AandA...699A.342B} \\
\hline
\end{tabular}

\tablefoot{
(a) The stellar abundances were adopted from the Hypatia catalogue \citep{2014AJ....148...54H}. The dust-to-gas ratio was scaled with the stellar metallicity for each system.
(b) The planetary C/H, O/H, and C/O values correspond to retrieved atmospheric abundances from the literature.
}
\end{table*}

\section{Tidal circularisation and realignment timescales}
\label{appendixD}

The tidal circularisation timescale is estimated using the equilibrium tide model \citep{2008ApJ...678.1396J}:
\begin{equation}
\tau_{\rm circ} =
\frac{4}{63}
\frac{Q_p}{k_{2p}}
\frac{M_p}{M_\star}
\left(\frac{a}{R_p}\right)^5
\frac{P_{\rm orb}}{2\pi},
\end{equation}
where $Q_p$ is the planetary tidal quality factor, $k_{2p}$ is the planetary tidal Love number, $M_p$ and $M_\star$ are the planetary and stellar masses, $a$ is the orbital semi-major axis, $R_p$ is the planetary radius, and $P_{\rm orb}$ is the orbital period.
The spin--orbit realignment timescale is estimated following \citet{2012MNRAS.423..486L}:
\begin{eqnarray}
\tau_{\rm realign} = 4.3\ \text{Gyr}
\left(\frac{\kappa}{0.1}\right)
\left(\frac{Q_{10}/k_{10}}{10^{7}}\right)
\left(\frac{M_\star}{10^{3} M_{p}}\right)^{2}
\left(\frac{\overline{\rho}_{*}}{\overline{\rho}_{\odot}}\right)  
\left(\frac{10\ \text{d}}{P_{s}}\right)
\left(\frac{P_{\rm orb}}{1\ \text{d}}\right)^{4},
\end{eqnarray}
where $\kappa$ is the stellar moment of inertia coefficient, $Q_{10}/k_{10}$ is the effective tidal quality factor for inertial-wave dissipation, $\overline{\rho}_*$ is the mean stellar density, and $P_s$ is the stellar rotation period.

For these order-of-magnitude estimates, we adopted fiducial values of $Q_p/k_{2p}=10^6$, $\kappa=0.1$, and $Q_{10}/k_{10}=10^6$--$10^7$, following the normalisations commonly used in these tidal prescriptions \citep{2008ApJ...678.1396J,2012MNRAS.423..486L}. Because these tidal parameters are poorly constrained and model-dependent, the resulting timescales should be interpreted as indicative rather than precise.
For systems without directly measured stellar rotation periods, $P_s$ was estimated from $v\sin i_*$ and $R_*$. The obliquity transition near $T_{\rm eff}\simeq6250\,{\rm K}$ reflects the reduced efficiency of tidal realignment in hotter stars with thinner convective envelopes \citep{2010ApJ...718L.145W}.

\end{appendix}
\end{document}